\documentclass[useAMS,usenatbib]{mn2e}
\usepackage[dvips]{graphicx}
\usepackage{amssymb,lscape}

\title[Radio galaxies and their magnetic fields]{Radio galaxies and their magnetic fields out to $z\leq3$}
\author[J.~K.~Banfield et al.]{J.~K.~Banfield$^{1,2}$, D.~H.~F.~M.~Schnitzeler$^{3}$,  S.~J.~George$^{4}$, R.~P.~Norris$^{1}$,\newauthor T.~H.~Jarrett$^{5}$, A.~R.~Taylor$^{5,6}$, J.~M.~Stil$^{7}$\\
$^{1}$CSIRO Australia Telescope National Facility, PO Box 76, Epping, NSW, 1710, Australia\\
$^{2}$Research School of Astronomy and Astrophysics, Australian National University, Weston Creek, ACT 2611, Australia\\
$^{3}$Max Planck Institut fur Radioastronomie, 53121 Bonn, Germany\\
$^{4}$Astrophysics and Space Research Group, School of Physics and Astronomy, University of Birmingham, Birmingham, B15 2TT, UK\\
$^{5}$Astronomy Department, University of Cape Town, Rondebosch 7701, Republic of South Africa\\
$^{6}$Department of Physics, University of the Western Cape, Bellville 7535, Republic of South Africa\\
$^{7}$Department of Physics and Astronomy, The University of Calgary, 2500 University Drive NW, Calgary AB, T2N 1N4, Canada}

\begin{document}

\date{Accepted 2014 July 11.  Received 2014 July 8; in original form 2013 November 14}

\pagerange{\pageref{firstpage}--\pageref{lastpage}} \pubyear{2014}

\maketitle

\label{firstpage}

%-------------------------------------------------------------------------
%                               ABSTRACT
%-------------------------------------------------------------------------
\begin{abstract}
We present polarisation properties at $1.4\,$GHz of two separate extragalactic source populations: passive quiescent galaxies and luminous quasar-like galaxies.  We use data from the {\it Wide-Field Infrared Survey Explorer} data to determine the host galaxy population of the polarised extragalactic radio sources.  The quiescent galaxies have higher percentage polarisation, smaller radio linear size, and $1.4\,$GHz luminosity of $6\times10^{21}<L_{\rm 1.4}<7\times10^{25}\,$W Hz$^{-1}$, while the quasar-like galaxies have smaller percentage polarisation, larger radio linear size at radio wavelengths, and a $1.4\,$GHz luminosity of $9\times10^{23}<L_{\rm 1.4}<7\times10^{28}\,$W Hz$^{-1}$, suggesting that the environment of the quasar-like galaxies is responsible for the lower percentage polarisation.  Our results confirm previous studies that found an inverse correlation between percentage polarisation and total flux density at $1.4\,$GHz.   We suggest that the population change between the polarised extragalactic radio sources is the origin of this inverse correlation and suggest a cosmic evolution of the space density of quiescent galaxies.  Finally, we find that the extragalactic contributions to the rotation measures (RMs) of the nearby passive galaxies and the distant quasar-like galaxies are different. After accounting for the RM contributions by cosmological large-scale structure and intervening Mg\,{II} absorbers we show that the distribution of intrinsic RMs of the distant quasar-like sources is at most four times as wide as the RM distribution of the nearby quiescent galaxies, if the distribution of intrinsic RMs of the WISE-Star sources itself is at least several rad m$^{-2}$ wide.
\end{abstract}


%-------------------------------------------------------------------------
%                             INTRODUCTION
%-------------------------------------------------------------------------
\section{Introduction}
\citet{Tucci2012} recently found that the intrinsic percentage polarisation of extragalactic radio sources (ERS) at frequencies $\ge 20\,$GHz is between $2-5\,$per cent, independent of flux density.  These results were confirmed by \citet{Massardi2013} using the Australia Telescope $20\,$GHz (AT20G) Survey, while \citet{Sadler2006} suggest that there is a trend that fainter sources tend to have higher percentage polarisation.  This anti-correlation between percentage polarisation and total flux density has also been suggested at $1.4\,$GHz by \citet{Mesa2002}, \citet{Tucci2004}, \citet{Taylor2007}, \citet{Subrahmanyan2010}, and \citet{Grant2010}.  Recently, \citet{Hales2013} found no evidence for this trend and attribute the previous results to selection effects consistent with the reasoning by \citet{Massardi2013}.  As a result the anti-correlation of percentage polarisation with total flux density, if it exists, remains a mystery.

The first studies of increasing percentage polarisation with decreasing flux density came from \citet{Mesa2002} and \citet{Tucci2004} who both suggested a population change of ERS at fainter flux densities was the cause.  \citet{Taylor2007} went on to suggest that the cause was a result of a change in the fraction of radio quiet active galactic nuclei (AGN).  Most recently, \citet{Rudnick2014} examined the polarisation properties of radio sources down to $S_{\rm 1.4\,GHz} > 15\,\mu$Jy in the GOODS-N field and suggest a population change around a polarised flux density of $1\,$mJy.  Studies into the intrinsic properties of polarised ERS by \citet{Banfield2011} show a trend of increasing percentage polarisation with decreasing luminosity and no trend with redshift, later confirmed by \citet{Hammond2012}.  \citet{Subrahmanyan2010} suggest that this anti-correlation between percentage polarisation and total flux density is likely to be a transition from FRII-dominated to FRI-dominated populations, while the results by \citet{Grant2010} imply that the higher percentage polarisation may be originating in the lobe-dominated structure and not in beamed BL Lac objects.  However, \citet{Shi2010} found no dependence on ERS environment when comparing highly polarised ($> 30\,$per cent) ERS with their low polarised counterparts.  \citet{Shi2010} went on to suggest that intrinsic properties of magnetic field ordering, thermal plasma density, and magnetic field orientation to the line of sight are the root cause for highly polarised ERS.

In this paper we present an analysis of 1.4~GHz polarised ERS in combination with optical spectroscopic data in order to explore this anti-correlation between percentage polarisation and flux density.   We probe polarised radio emission out to high-redshifts and examine the magnetic fields within different ERS populations.  We outline the sample selection in Section \ref{sec:sample} and the nature of the polarised sources is discussed in Section \ref{sec:nature}.  Section \ref{sec:seleff} describes the selection effects of our data, we discuss our findings in Section \ref{sec:dis}, and conclusions are presented in Section \ref{sec:conc}. 

The cosmological parameters used throughout this paper are: $\Omega_{\rm \lambda} = 0.7$; $\Omega_{\rm M} = 0.3$; and $H_{0} = 70$~kms$^{-1}$Mpc$^{-1}$.   We define the spectral index $\alpha$ as $S\propto \nu^{\alpha}$.

%-------------------------------------------------------------------------
%                           MAGNETIC FIELDS
%-------------------------------------------------------------------------
\section{Observational Indications of Cosmic Magnetic Fields}
%%% Polarisation %%%
\subsection{Measuring Polarisation}
All extragalactic radio sources emit radiation that is partially polarised and a measurement of all four Stokes parameters provides the necessary information to completely describe the polarisation state of the electromagnetic radiation received from a radio source. Stokes $I$ represents the total amount of radiation received, Stokes $Q$ and Stokes $U$ contain the linearly polarised information, while Stokes $V$ contains the circularly polarised information. The linearly polarised flux density of a radio source is calculated by:
\begin{equation}
P = \sqrt{Q^2+U^2} \, ,
\end{equation}
and the percentage polarisation is calculated by:
\begin{equation}
\Pi=\left(\frac{\rm total \, polarised \, flux}{\rm total \, flux}\right) \times 100\% = \frac{P}{S} \times 100\% \, .
\end{equation}

The statistical distribution of the noise when measuring polarised intensity is non-Gaussian and has a non-zero mean.  Therefore the resulting value of $P$ can be biased high depending on the signal-to-noise.  The removal of this bias can be estimated as $P_{\rm 0} = \sqrt{P^2 - \sigma^2}$ for a signal-to-noise greater than 4 \citep{Simmons1985}.  Also, errors derived from the least-squares approach will be too small \citep{Wardle1974}.

%%% RM %%%
\subsection{Faraday Rotation}
The amount of Faraday rotation of polarized radio waves provides information on the strength and structure of the magnetic field along the line of sight, and  depends on three factors: (1) the wavelength of the emission; (2) the electron density of the medium; and (3) the strength of the line-of-sight component of the magnetic field in the medium. 
Expressed in equation form:
\begin{equation}
\Phi = \lambda^2 \left(0.812\int n_\mathrm{e} \vec{B}\cdot d\vec{l}\right) \, {\rm rad} \, ,
\end{equation} 
where $\vec{B}$ is the magnetic field ($\mu \mathrm{G}$), $d\vec{l}$ is an infinitesimal distance along the line of sight towards the observer (pc), $\lambda$ is the observing wavelength (m), and $n_\mathrm{e}$ is the electron density (cm$^{-3}$).  The rotation measure (RM) is given by:
\begin{equation}
\mathrm{RM} = 0.812\int n_\mathrm{e} \vec{B}\cdot d\vec{l} \,\,\,\, {\rm rad \, m^{-2}} \, ,
\end{equation}
and is integrated from the source of the polarized radio waves to the observer. A positive RM indicates that the magnetic field component along the line of sight points towards the observer, while a magnetic field pointing away from the observer produces a negative RM.

Many factors contribute to the observed RMs of extragalactic radio sources, such as the Earth's ionosphere (RM$_{\rm ion}$), the Milky Way foreground (RM$_{\rm MW}$), and any Faraday screens that could be intrinsic to the sources or lie between the source of the emission and the Milky Way, which we shall we refer to as the `extragalactic RM' (RM$_{\rm ERS}$).  The sum of all these contributions is the RM that is measured:
\begin{equation}
\mathrm{RM} = \mathrm{RM}_\mathrm{ion}+\mathrm{RM}_\mathrm{MW} + \mathrm{RM}_\mathrm{ERS} \, ,
\end{equation}
where RM$_{\rm ion}$ is typically 1 to $2\,$rad m$^{-2}$ \citep{Sotomayor2013}.  
In section \ref{sec:rm}, we derive a new method for extracting the RM$_{\rm ERS}$ for different radio source populations.

%-------------------------------------------------------------------------
%                               DATA SAMPLE
%-------------------------------------------------------------------------
\section{Sample Selection}\label{sec:sample}
%%% RM catalogue %%%
\subsection{Rotation Measure and Redshift Catalogue}\label{sec:RMcat}
We used data from the \citet{Hammond2012} catalogue, which contain spectroscopic redshifts for 4003 polarised radio sources from the rotation measure catalog of \citet{Taylor2009} at $1.4\,$GHz with a declination $\delta \ge -40^{\circ}$ and a flux density $S_{\rm 1.4\,GHz}\ge 11\,$mJy in the redshift range $0<z<5.3$.  The polarisation information comes from the NRAO VLA Sky Survey \citep[NVSS;][]{Condon1998} which has an angular resolution of $45''$ and includes only those sources with a signal-to-noise greater than 8 so that the noise bias correction for polarised flux density is negligible \citep{Simmons1985,George2012}.  \citet{Hammond2012} extracted redshifts from optical counterparts in the NASA/IPAC Extragalactic Database \citep[NED;][]{Helou1991}, Set of Identifications, Measurements and Bibliography for Astronomical Data \citep[SIMBAD;][]{Wenger2000}, Sloan Digital Sky Survey \citep[SDSS;][]{York2000}, Six-degree Field Galaxy Survey \citep[6dFGS;][]{Jones2009}, Two-degree Field Galaxy Redshift Survey \citep[2dFGRS;][]{Colless2001}, and the 2dF QSO Redshift survey/6dF QSO Redshift survey \citep[2QZ/6QZ;][]{Croom2004}.  
%%% WISE %%%
\subsection{Wide-field Infrared Survey Explorer}
\begin{table}
 \centering
  \caption{Distribution of polarised NVSS sources from the \citet{Hammond2012} catalogue with a $5\sigma$ detection in each WISE band.}\label{wisetable}
 \begin{tabular}{lrr}
 \hline
WISE Band & N & Fraction of \\
 & &  sources detected\\
\hline
$3.4\,\mu$m & 3741 & 93.5\,$\pm$\,1.5\%\\
$4.6\,\mu$m & 3693 & 92.3\,$\pm$\,1.5\%\\
$12\,\mu$m & 2729 & 68.2\,$\pm$\,1.3\%\\
$22\,\mu$m & 1440 & 40.0\,$\pm$\,1.1\%\\
\hline
Any Band & 3747 & 93.6\,$\pm$\,1.5\%\\
\hline
\end{tabular}
\end{table}
The {\it Wide-field Infrared Survey Explorer} \citep[WISE;][]{Wright2010} surveyed the sky at wavelengths $3.4$, $4.6$, $12$, and $22\,\mu$m with a $5\sigma$ point source sensitivity in unconfused regions of at least $0.08$, $0.11$, $1.0$, and $6.0\,$mJy and angular resolutions of $6.1''$, $6.4''$, $6.5''$, and $12.0''$.  This sensitivity depends on the ecliptic latitude, with the poles having the greatest depth \citep{Jarrett2011}.  The selection of these four bands makes WISE an excellent instrument for studies of stellar structure and interstellar processes of galaxies.  The two shorter bands trace the stellar mass distribution in galaxies and the longer wavelengths map the warm dust emission and polycyclic aromatic hydrocarbon (PAH) emission, both tracing the current star formation activity.

Using the optical counterparts from the \citet{Hammond2012} catalogue, we matched 3747 polarised radio sources to within $5\arcsec$ of their WISE ALLSky Source Catalogue\footnote{http://irsa.ipac.caltech.edu/Missions/wise.html} counterparts down to a $5\sigma$ detection in at least one of the four WISE bands.  Table \ref{wisetable} lists the distribution of sources in the four WISE bands.   

%-------------------------------------------------------------------------
%                                  NATURE
%-------------------------------------------------------------------------
\section{Nature of Polarised Radio Sources}\label{sec:nature}
% Tab. 2
\begin{table*}
\centering
\caption{Summary of key results from section \ref{sec:nature} for the two WISE populations of polarised ERS including: redshift range, median redshift, spectral index range, median spectral index, luminosity range, median angular and linear sizes, and rotation measure. The errors in $\sigma_\mathrm{err}$ indicate the 1-sigma range.}\label{tab:nature}
\begin{tabular}{lccccc}
\hline
Population & $z$ & $<z>$ & $\alpha_{325}^{1400}$ & $<\alpha_{325}^{1400}>$ & $L_{\rm 1.4}$ (W Hz$^{-1}$)\\
\hline
Full & $0.001<z<5.3$ & 0.78 & $-2.62<\alpha< 3.36$ & $-0.68\,\pm\,0.02$ & $ 6\times10^{21}<L< 1\times10^{29}$ \\
WISE--AGN & $0.500<z<3.7$ & 1.03 &  $-2.62<\alpha< 3.01$ & $-0.63\,\pm\,0.02$ & $  9\times10^{23}<L<  7\times10^{28}$ \\
WISE--Star & $0.006<z<0.8$ & 0.06 & $-1.94<\alpha< 1.50$ & $-0.59\,\pm\,0.05$ & $  6\times10^{21}<L<  7\times10^{25}$ \\
\hline
Population & $<\theta_{\rm AS}>$ (\arcsec) & $<\theta_{\rm LS}>$ (kpc) & $\sigma_{\rm ERS}$ (rad m$^{-2}$) & $\sigma_{\rm err}$ (rad m$^{-2}$)\\
\hline
Full & $37\,\pm\,1$ &$214\,\pm\,5$ & $\,\,\,6.3 < \sigma_{\rm ERS} < \,\,\,6.9$ ($\pm > 0.3$) & $10.3\,\pm\,0.1$\\
WISE--AGN & $34\,\pm\,1$ & $253\,\pm\,8$ & $12.0 < \sigma_{\rm ERS} < 12.1$ ($\pm > 0.2$) & $6.4 < \sigma_{\rm err} < 6.5\,\pm\,0.2$\\
WISE--Star & $58\,\pm\,5$ & $\,\,\,69\,\pm\,8$ & $\,\,\,7.4 < \sigma_{\rm ERS} < 8.7$ ($\pm > 1.0$) & $9.2 < \sigma_{\rm err} < 9.4\,\pm\,0.9$\\
\hline
\end{tabular}
\end{table*}
The key findings of this section are summarised in Table \ref{tab:nature}.

%%% WISE Colours %%%
\subsection{WISE Colours of Polarised ERS}\label{sec:wise}
% Fig. 1
\begin{figure}
\includegraphics[scale=0.40]{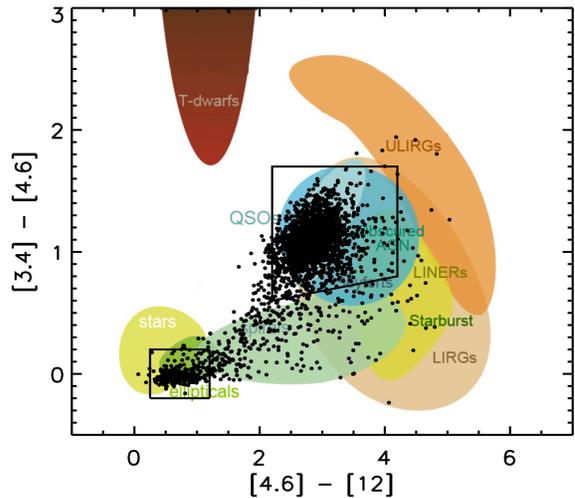}
\caption{WISE colour-colour diagram, plotted in units of magnitude, for the polarised ERS with a $5\sigma$ detection in the three WISE bands of 3.4, 4.6, and $12\,\mu$m.  The upper-right box indicates the region of WISE--AGN and the lower-left box indicates the region of WISE--Stars; all areas are described by \citet{Jarrett2011}.}
\label{colormag}
\end{figure}
% Fig. 2
\begin{figure}
\begin{center}
\includegraphics[scale=0.40]{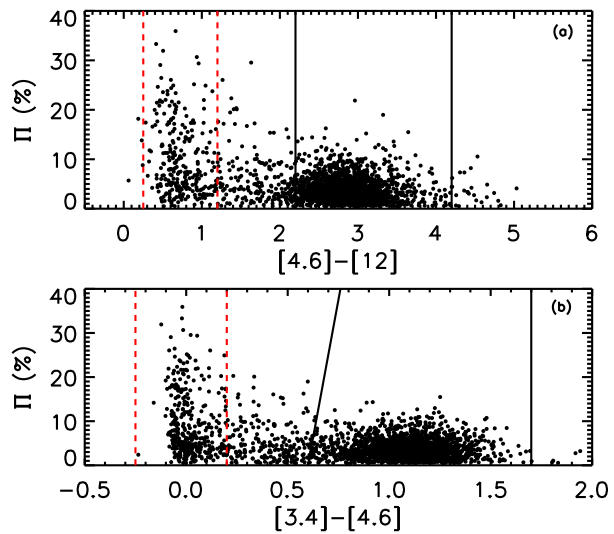}
\caption{Percentage polarisation as a function of WISE colour (a) $[4.6]-[12]$ and (b) $[3.4]-[4.6]$ from Fig.~\ref{colormag}.  The red dashed lines indicate the boundary of WISE--Star region and the solid black lines indicate the boundary of the WISE--AGN region as defined in Fig.~\ref{colormag}.}
\label{colormag-pi}
\end{center}
\end{figure}
%Fig. 3
\begin{figure*}
\begin{center}
\includegraphics[scale=0.43]{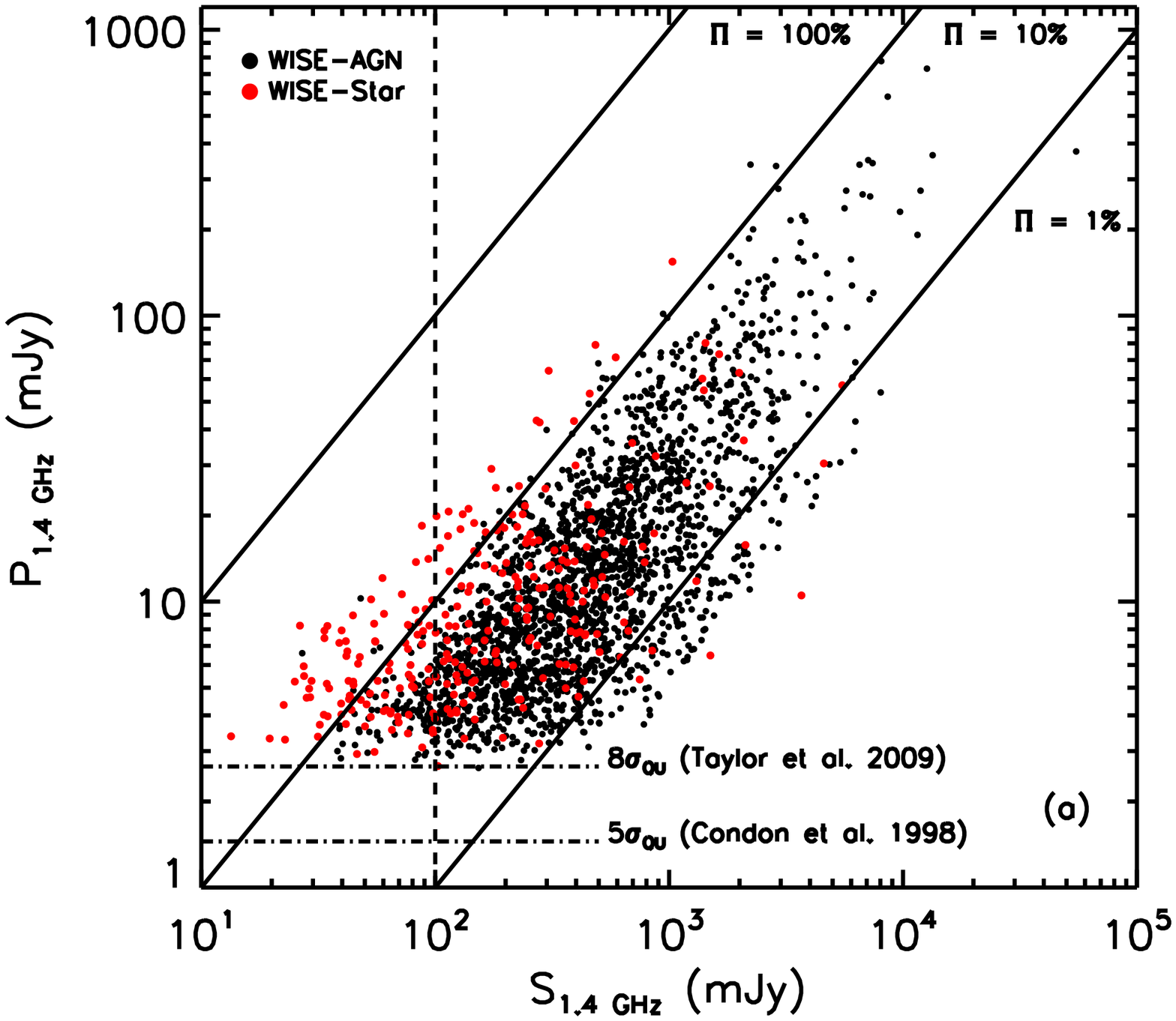}
\includegraphics[scale=0.43]{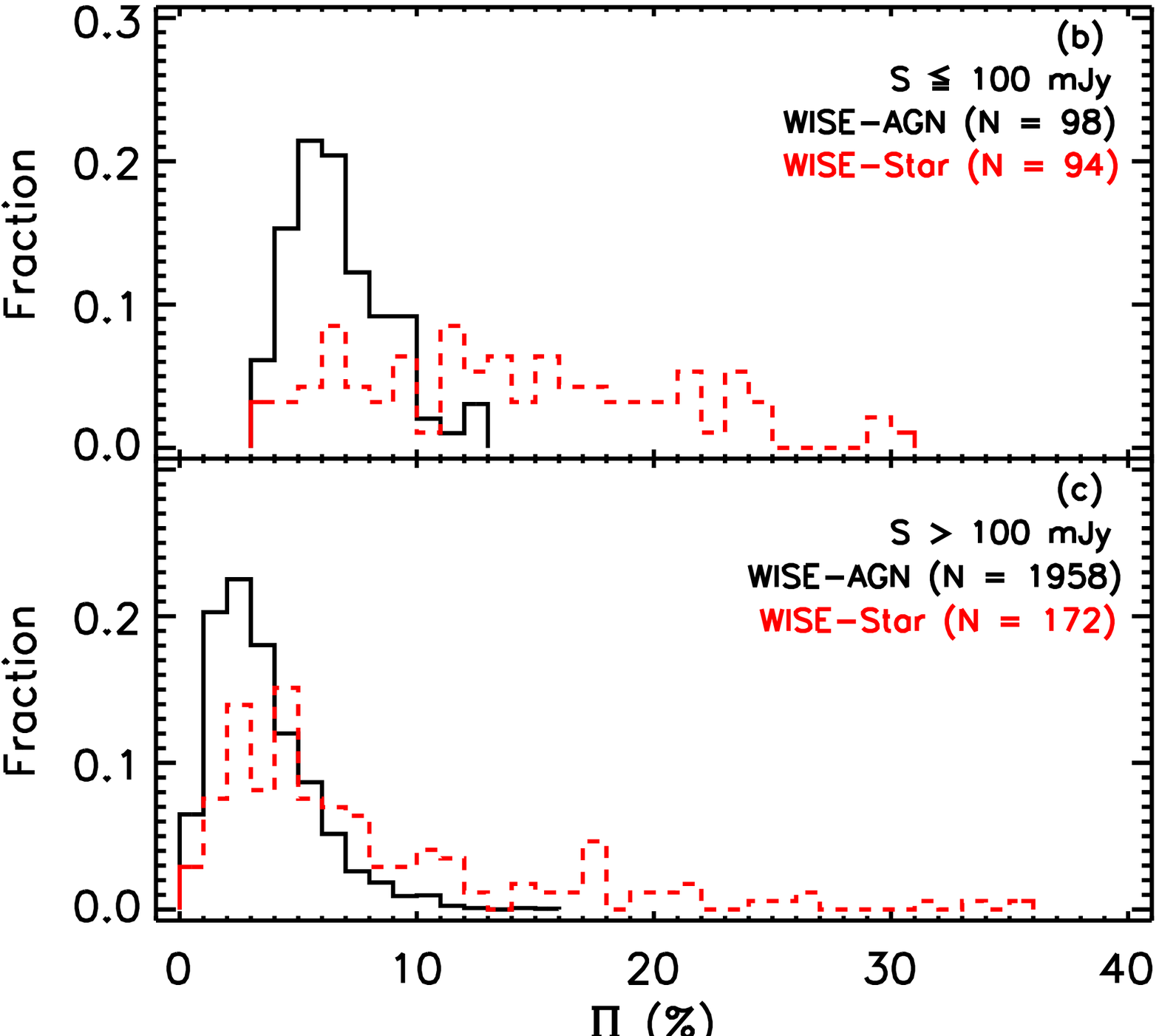}
\caption{(a) Distribution of log($\,S_{\rm 1.4\,GHz}$) vs.~log($\,P_{\rm 1.4\,GHz}$) for the polarised ERS in Fig.~\ref{colormag}.  The red dots indicate the polarised ERS in the WISE--Star region, while the black dots indicate the polarised ERS in the WISE--AGN region.  The solid diagonal lines indicate the $\Pi=1$ (right), 10 (middle), and 100 per cent (left) percentage polarisation levels. Also plotted are the $8\sigma_{QU}$ flux density limit of the \citet{Taylor2009} catalogue and the $5\sigma_{QU}$ flux density limit from NVSS \citep{Condon1998}.  (b) Percentage polarisation histogram of the fraction of polarised ERS in the WISE--AGN (black line) and WISE--Star (red line) regions for $S_{\rm 1.4\,GHz} \leq 100\,$mJy as indicated by the vertical dashed line in (a).  (c) The same histogram as in (b) but for $S_{\rm 1.4\,GHz} > 100\,$mJy.}
\label{logIlogP}
\end{center}
\end{figure*}

The WISE $[4.6]-[12]$ and $[3.4]-[4.6]$ colour-colour diagram is a good tool to distinguish different galaxy populations as outlined by the coloured shapes in Fig.~\ref{colormag}, which have been defined by \citet{Wright2010} and \citet{Jarrett2011}.  Infrared emission dominated by light from evolved stars is found near zero colour, stretching to redder colours along the $[4.6]-[12]$ axis towards more luminous evolved populations as traced by the $12\,\mu$m light and the power-law mid-infrared spectrum of AGN dominates the redder WISE colours in $[3.4]-[4.6]$ \citep{Jarrett2011,Stern2012}.  The general ``AGN'' region covering QSOs to Seyfert galaxies, as defined by \citet{Jarrett2011}, is illustrated by the box shown in upper right of Fig.~\ref{colormag} (hereafter WISE--AGN).  The region redward of the AGN box is populated by the most luminous infrared galaxies, while the region defined to be dominated by the infrared emission from starlight is defined by the box shown in the lower left of Fig.~\ref{colormag} (hereafter WISE--Star).  The dots in Fig.~\ref{colormag} show the $2724$ polarised NVSS ERS with a $5\sigma$ WISE detection in the $3.4$, $4.6$, and $12\,\mu$m bands.  The polarised ERS in our sample clearly fall primarily in these two regions, with $266$ polarised ERS in the region defined as WISE--Star and $2056$ ERS in the WISE--AGN region.  

In Fig.~\ref{colormag-pi} we plot the percentage polarisation of the sources from Fig.~\ref{colormag} as a function of the WISE colours.  Fig.~\ref{colormag-pi}(a) shows the $[4.6]-[12]$ WISE colour and Fig.~\ref{colormag-pi}(b) shows the $[3.4]-[4.6]$ WISE colour along with the boundary of the WISE--AGN (black lines) and WISE--Star (red dashed lines) regions. The mean percentage polarisation of the WISE--AGN ERS $<\Pi_{\rm AGN}>=3.6\pm0.2\,$per cent (median $\Pi_{\rm AGN}=3.2\,$per cent) and for the WISE--Star ERS $<\Pi_{\rm Star}>=10.0\pm0.5\,$per cent (median $\Pi_{\rm Star}=7.5\,$per cent).  The polarised ERS that fall into the WISE--Star region show a higher percentage polarisation by a factor of 3 than the polarised ERS in the WISE--AGN region.

The distribution of total flux density ($S_{\rm 1.4\,GHz}$) and polarised flux density ($P_{\rm 1.4\,GHz}$) can be seen in Fig.~\ref{logIlogP}(a).  The polarised WISE--AGN ERS are indicated by black dots, while the polarised WISE--Star ERS are indicated by red dots; the solid diagonal lines indicate $\Pi=1, 10,$ and 100 per cent.  We split the sample into two flux density bins: (1) $S_{\rm 1.4 GHz} > 100\,$mJy; and (2) $S_{\rm 1.4 GHz} \le 100\,$mJy which results in a similar number of sources with $S_{\rm 1.4 GHz} \le 100\,$mJy for both polarised ERS populations.  We plot the percentage polarisation distribution of the two polarised ERS populations in Fig.~\ref{logIlogP}(b) for $S_{\rm 1.4 GHz} \le 100\,$mJy and Fig.~\ref{logIlogP}(c) for $S_{\rm 1.4 GHz} >100\,$mJy.

For the polarised ERS with $S_{\rm 1.4 GHz} \le 100\,$mJy, Fig.~\ref{logIlogP}(b), we have 98 WISE--AGN sources and 94 WISE--Star sources.  We found the mean percentage polarisation $<\Pi_{\rm AGN}>=6.7\pm0.2\,$per cent for the WISE--AGN population and $<\Pi_{\rm Star}>=14.0\pm0.7\,$per cent for our WISE--Star population.  We did a KS test of the percentage polarisation of the two ERS populations and found that the two distributions are not drawn from the same parent population at the 99 per cent significance level. 

For the polarised ERS with $S_{\rm 1.4 GHz} > 100\,$mJy, Fig.~\ref{logIlogP}(c), we have 1958 WISE--AGN sources and 172 WISE--Star sources. We found the mean percentage polarisation $<\Pi_{\rm AGN}>=3.5\pm0.1\,$per cent for the WISE--AGN population and $<\Pi_{\rm Star}>=7.7\pm0.5\,$per cent for our WISE--Star population.  We did a KS test of the percentage polarisation of the two ERS populations and found that the two distributions are not drawn from the same parent population at the 99 per cent significance level.

%%% REDSHIFT %%%
\subsection{Redshift Distribution}\label{sec:redshift}
% Fig. 4
\begin{figure}
\begin{center}
\includegraphics[scale=0.40]{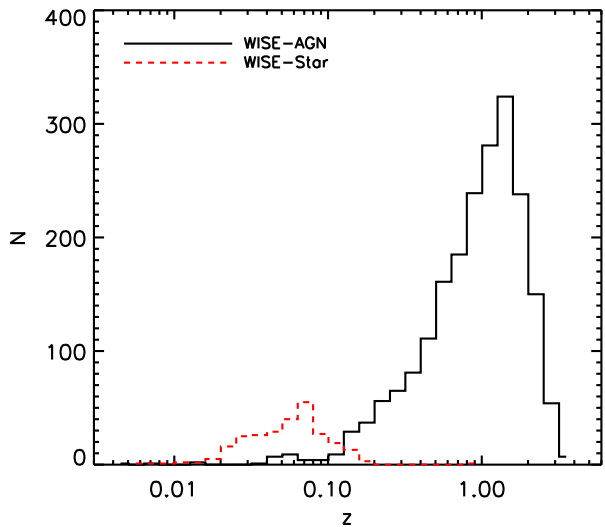}
\caption{Redshift distribution of polarised ERS in the two regions defined in Fig.~\ref{colormag}.  The red dashed line represents the 266 polarised WISE--Star ERS.  The black solid line represents the 2056 polarised WISE--AGN ERS.}
\label{redshiftdist}
\end{center}
\end{figure}

Fig.~\ref{redshiftdist} shows the redshift distribution of the two source populations as defined in Section \ref{sec:wise}.   The polarised WISE--Star population are found to be low-redshift galaxies in the range $0.006 < z < 0.8$, while the polarised WISE--AGN population are high redshift galaxies, and therefore more luminous, in the range $0.5 < z < 3.7$.  We recognise that WISE is not sensitive to early-type galaxies at high redshift as they are too faint.  Our sample of galaxies at high redshift must consist of luminous AGN or starburst or a mixed population of both.
%%% LUMINOSITY %%%
\subsection{{\rm $1.4\,$GHz} Monochromatic Luminosity Distribution}\label{sec:catalogue}
% Fig. 5
\begin{figure}
\begin{center}
\includegraphics[scale=0.40]{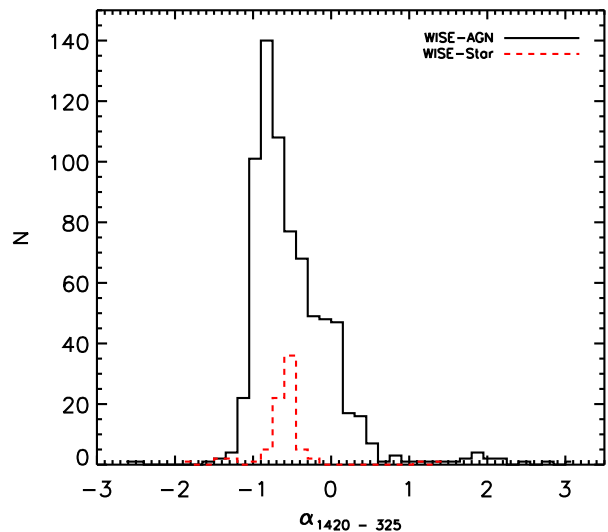}
\caption{Spectral index distribution between $1400 - 325\,$MHz of the polarised ERS in our sample.  The WISE--Star ERS is indicated by the red dashed line and the WISE--AGN ERS is given by the black solid line.  The median spectral index for the WISE--Star population is $-0.59\,\pm\,0.05$ and $-0.63\,\pm\,0.02$ for the WISE--AGN polarised ERS population.}
\label{sindex}
\end{center}
\end{figure}
% Fig. 6
\begin{figure}
\begin{center}
\includegraphics[scale=0.40]{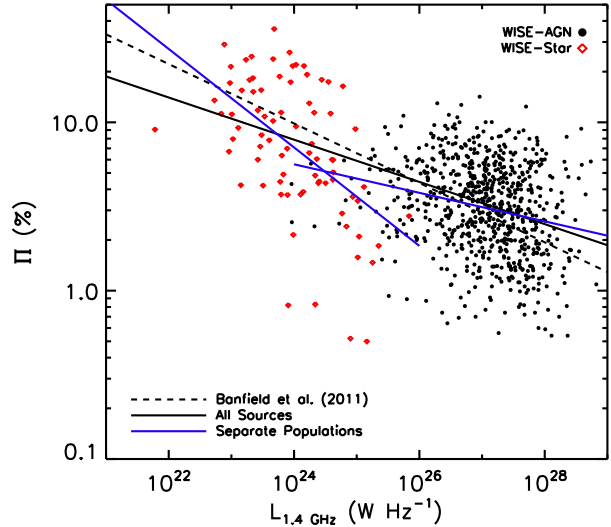}
\caption{Percentage polarisation versus luminosity for our sample of polarised ERS.  The polarised WISE--AGN ERS are shown by the black dots, while the WISE--Star ERS are shown with red diamonds.  A power law was fit to the full data set (solid black line), only to the WISE--AGN population (solid blue line on right), and only to the WISE--Star population (solid blue line on left).  For comparison, the fit from \citet{Banfield2011} is shown as the black dashed line.  For the full sample of polarised ERS $\beta = -0.13\,\pm\,0.01$, for the WISE--AGN population $\beta = -0.08\,\pm\,0.01$, and $\beta = -0.29\,\pm\,0.05$ for the WISE--Star polarised ERS population.}
\label{lum}
\end{center}
\end{figure}

The $1.4\,$GHz monochromatic luminosity ($L_{\rm1.4\,GHz}$) of the $4003$ radio sources from the \citet{Hammond2012} catalog was calculated using the equations of \cite{Hogg1999}:
\begin{equation}
L_{\rm1.4\,GHz} = \frac{4\pi D_{L}^{2} S_{\rm1.4\,GHz}}{(1+z)}(1+z)^{-\alpha} \, ,
\end{equation}
where $D_{L}$ is the luminosity distance, $S_{\rm 1.4\,GHz}$ is the flux density at $1.4\,$GHz, $\alpha$ the spectral index and $z$ is the redshift.  

Spectral indices are required to calculate the monochromatic luminosity of the ERS.   We calculated spectral indices from the $325\,$MHz Westerbork Northern Sky Survey \citep[WENSS;][]{Rengelink1997}, which covers declinations $\ge 28.5^{\circ}$.  WENSS has an angular resolution of $54''\times\,54''\mathrm{cosec}{\delta}$ and contains more than $200,000$ sources down to a flux density of $S_{\rm 325\,MHz} = 18\,$mJy.  Fig.~\ref{sindex} shows the distribution of spectral indices for the two WISE populations of polarised ERS.  The median spectral index of polarised WISE--Star ERS (N$_{\rm Star}=78$) is $-0.59\,\pm\,0.05$, while the polarised WISE--AGN ERS (N$_{\rm AGN}=731$) is $-0.63\,\pm\,0.02$.

In Fig.~\ref{lum}, we plot the monochromatic luminosity of polarised ERS in the two WISE populations.   The luminosity range of our full sample of polarised ERS is $6\times10^{21} < L_{\rm 1.4 GHz} < 1\times10^{29}\,$W Hz$^{-1}$, while for the WISE--AGN polarised ERS population the range is $9\times10^{23} < L_{\rm 1.4 GHz} < 7\times10^{28}\,$W Hz$^{-1}$, and WISE--Star galaxies have $6\times10^{21} < L_{\rm 1.4 GHz} < 7\times10^{25}\,$W Hz$^{-1}$.   Our two WISE polarised ERS populations split into two separate regions around $L_{\rm 1.4 GHz}\sim10^{25}\,$W Hz$^{-1}$ with the WISE--Star population filling out the lower-luminosity side of the plot and the WISE--AGN population filling out the higher-luminosity side of the plot.  Following the relationship from \citet{Banfield2011}, a power law of the form:
\begin{equation}
\frac{\Pi}{\Pi_{\rm 0}} = \left( \frac{L_\nu}{L_{\rm 0}} \right)^\beta \, ,
\end{equation}
was fit to the data.  For the full sample of polarised ERS $\beta = -0.13\,\pm\,0.01$, for the WISE--AGN population $\beta = -0.08\,\pm\,0.01$ and $\beta = -0.29\,\pm\,0.05$ for the WISE--Star population.  We ran a Spearman rank correlation test on the two population fits to determine the relationship between percentage polarisation and monochromatic luminosity at $1.4\,$GHz. There is a moderate negative linear correlation for the WISE--Star population ($r_{s,{\rm star}}=-0.53\,\pm\,0.07, N_{\rm star}=78, p < 0.01$) and a weak negative linear correlation for the WISE--AGN population ($r_{s,{\rm AGN}}=-0.24\,\pm\,0.02, N_{\rm AGN}=731, p < 0.01$).

%%% RM SECTION %%%
\subsection{Rotation Measures}\label{sec:rm}
In order to compare the extragalactic contributions to the RMs of the nearby WISE--Star sources and the distant WISE--AGN we have to correct for the RM contributions by the Galactic foreground and measurement errors. We follow the method described in \citet{Schnitzeler2010} to correct for Galactic RM foregrounds, which separates Galactic from extragalactic RM contributions based on the idea that the former contributions are correlated between sightlines, while the latter are not. First we split the data into strips along Galactic longitude and use cubic spline fitting to remove large-scale RM gradients along Galactic longitude. The strips span only a narrow range in Galactic latitude to suppress the variation in Galactic RM with Galactic latitude. 
In Appendix~\ref{sec:appendixA} we show how the measured RM variance of the ensemble after the cubic spline fitting can be written in terms of the RM variance that is built up outside the Milky Way ($\sigma^2_{\rm ERS}$), the variance that is due to measurement errors ($\sigma^2_{\rm err}$), and a residual RM variance due to the Milky Way ($\sigma^2_{\rm MW}$) that could not be removed as:
\begin{eqnarray}\label{eqn:rm}
\sigma_\mathrm{RM}^2 & = & \left(\sigma_\mathrm{ERS}^2 + \sigma_\mathrm{err}^2\right)\left(\frac{N_\mathrm{los}-N_\mathrm{strips}}{N_\mathrm{los}-1}\right) + \nonumber \\
 & & \sum_{\mathrm{strip}\ i=1}^{N_\mathrm{strips}} \frac{N_i}{N_\mathrm{los}-1}\left(\langle\mathrm{RM}\rangle_{\mathrm{strip}\ i} - \langle\mathrm{RM}\rangle_{\mathrm{all\ strips}} \right)^2 + \nonumber \\
& & \sum_{\mathrm{strip}\ i=1}^{N_\mathrm{strips}}\left(\frac{N_i-1}{N_\mathrm{los}-1}\right)\sigma_{\mathrm{MW},i}^2 ,  
\end{eqnarray}
\noindent
where $N_\mathrm{los}$ is the total number of sightlines in the ensemble, $N_\mathrm{strips}$ is the total number of strips along Galactic longitude and $N_i$ is the number of useable sightlines in strip $i$.  
$\langle\mathrm{RM}\rangle_{\mathrm{strip}\ i}$ and $\langle\mathrm{RM}\rangle_{\mathrm{all\ strips}}$ indicate the mean RM in a single strip and the mean RM of all sightlines combined, respectively.
The strips in Equation~\ref{eqn:rm} are used to correct for the RM variance from the Milky Way that remains after cubic spline fitting ($\sigma_\mathrm{MW}^2$); these strips do not have to be the same as the strips that we used for cubic spline fitting. 
In our analysis we only use sightlines that lie further than 20$^\circ$ from the Galactic plane to avoid regions where the Galactic RM shows complex behaviour.
We consider only sightlines if they belong to strips with at least 5--15 sightlines (i.e., polarised ERS) in them; we vary this number, and we vary the width of the strips between 5--15 degrees in Galactic longitude to check how robust our results for $\sigma_\mathrm{ERS}^2$ are. If the number of sightlines in a strip is smaller than a specified minimum then all sightlines that belong to the strip are excluded from our analysis.
The mean, standard deviation, and variance are calculated using robust statistics that reject outliers at the 3-sigma level.

We test our method using all sightlines and find $\sigma_\mathrm{ERS} = 6.7-6.9\,$rad m$^{-2}$ ($\pm > 0.3$ rad m$^{-2}$). As we explain in Appendix~\ref{sec:appendixA}, we can calculate only a lower limit to the uncertainty in $\sigma_\mathrm{ERS}$. We then shift the strips by half a strip width to enable Nyquist sampling in Galactic latitude, and recalculate $\sigma^2_\mathrm{ERS}$, finding $\sigma_\mathrm{ERS} = 6.3-6.5\,$rad m$^{-2}$.  From a Monte Carlo simulation we derive $\sigma_\mathrm{err}$ = 10.3 $\pm$ 0.1 rad m$^{-2}$ (1--sigma). \citet{Schnitzeler2010} derived $\sigma_\mathrm{ERS}$ $\approx$ 6 rad m$^{-2}$ and $\sigma_\mathrm{err}$ = 10.4 $\pm$ 0.4 rad m$^{-2}$, in good agreement with the values we found. 

The polarised ERS identified as WISE--AGN have $\sigma_\mathrm{ERS}$ between 12.0 and 12.1 rad m$^{-2}$ ($\pm > 0.2$ rad m$^{-2}$), while $\sigma_\mathrm{err} = 6.4-6.5\,\pm\,0.2\,$rad m$^{-2}$. The extragalactic RM variance of the WISE--AGN is considerably larger than the extragalactic RM variance of the ensemble of all sources. The RM variance of the ensemble can be written as a weighted mean of the RM variances of the subpopulations; our observation that the subpopulation of WISE--AGN sources has a much larger RM variance than the ensemble as a whole implies that the subpopulations must cover a wide range of RM variances.
At different redshifts different subpopulations will contribute to the ensemble, which leads to a change in the RM variance in the ensemble as a function of redshift that could be misinterpreted as a signal from cosmological large-scale structure.
Therefore, when studying cosmological RM contributions, one should try to understand the composition of the ensemble of sources from which the RM variance is calculated.

Because the number of WISE--Star sources is so much smaller than the number of WISE--AGN sources, strips that we use to correct for $\sigma^2_{\rm MW}$ often contain fewer WISE--Star sources than the required minimum number of sources. We also found that for a strip width of 5 degrees the distribution of WISE--Star RMs could be non-Gaussian. To minimise the impact of  these two effects, for WISE--Star sources we only use strip widths of 10$^\circ$ and 15$^\circ$, and we found that $\sigma_\mathrm{ERS} = 7.4-8.7\,$rad m$^{-2}$ ($\pm > 1.0$ rad m$^{-2}$) and $\sigma_\mathrm{err}= 9.2 - 9.4\,\pm\,0.9\,$rad m$^{-2}$.  

Based on the $\sigma_\mathrm{ERS}$ of the WISE--Star and the WISE--AGN sources we conclude that they are different at the ($\lesssim$) 4-sigma level. In Appendix~\ref{sec:appendixA} we explain why the uncertainties in $\sigma_\mathrm{ERS}$ that we derive are lower limits, turning the statistical significance of the difference in $\sigma_\mathrm{ERS}$ between the two populations into an upper limit.

%%% SIZES %%%
\subsection{Polarised ERS Angular and Linear Size Distribution}\label{sec:size}
% Fig. 7
\begin{figure}
\begin{center}
\includegraphics[scale=0.2]{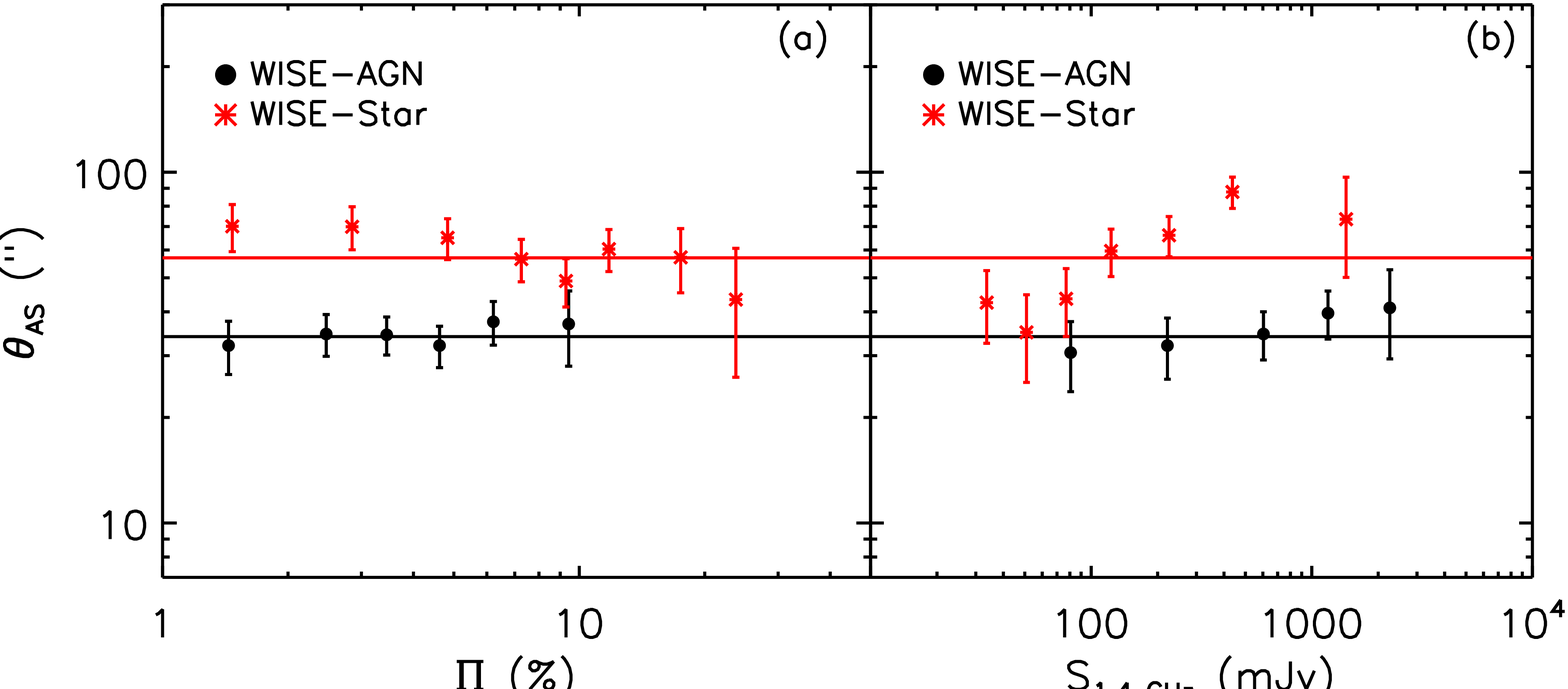}
\caption{(a) Distribution of median angular size as a function of percentage polarisation for the two WISE polarised ERS populations.  (b) Distribution of median angular size as a function of total flux density for the two WISE polarised ERS populations.  The WISE--AGN population is shown with black dots and the WISE--Star population is shown with red stars.  The errors bars are the standard error on the mean.  The median angular size of the two populations is shown with the solid lines.}
\label{angsizePI}
\end{center}
\end{figure}
% Fig. 8
\begin{figure}
\begin{center}
\includegraphics[scale=0.2]{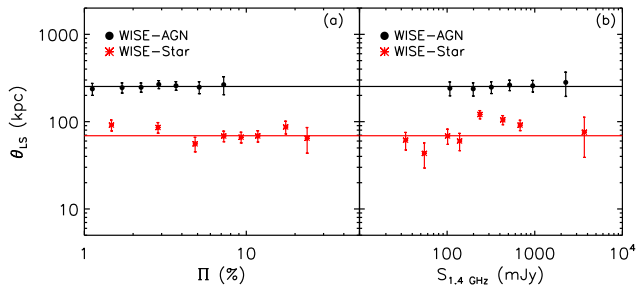}
\caption{(a) Distribution of median linear size as a function of percentage polarisation for the two WISE polarised ERS populations.  (b) Distribution of median linear size as a function of total flux density for the two WISE polarised ERS populations.  The WISE--AGN population is shown with black dots and the WISE--Star population is shown with red stars.  The errors bars are the standard error on the mean.  The median linear size of the two populations is shown with the solid lines.}
\label{Lsize}
\end{center}
\end{figure}

Using the Faint Images of the Radio Sky at Twenty Centimeters \citep[FIRST;][]{White1997} survey, we determined the angular size for 2088 of our polarised ERS.  We estimated the angular size by locating the $5\sigma$ boundary of the radio source and measuring the distance between boundary edges.  The median angular size of the polarised WISE--Star ERS population is $<\theta_{\rm AS}>= 58\,\pm\,1\arcsec$ and for the polarised WISE--AGN ERS population $<\theta_{\rm AS}>=34\,\pm\,1\arcsec$.  Both populations are resolved in FIRST ($\theta\sim5''$), consistent with polarised radio sources being resolved lobe-dominated sources \citep{Grant2010}.  Our findings are consistent with\citet{Rudnick2014} who found a small fraction of their polarised sources with angular sizes $>50\arcsec$.  In Fig.~\ref{angsizePI}, we plot the angular size distribution with total flux density and percentage polarisation distribution for the two separate WISE polarised ERS populations.  The polarised ERS were binned so that each bin contained roughly the same number of sources, and the median value of the angular size was determined for each bin.  The median values for the WISE--Star population are indicated by red stars in Fig.~\ref{angsizePI}, while the WISE--AGN population median angular size values are plotted with black dots.  There is a separation in the median angular sizes between the two WISE polarised ERS populations, with the WISE--Star population having larger angular sizes than the WISE--AGN population.  We note that the median angular sizes do not change significantly with percentage polarisation and total flux density for both populations.  

We investigated the effect of resolution by comparing the percentage polarisation of sources resolved in NVSS, $\theta_{\rm AS} \ge 45\arcsec$, and those unresolved in NVSS, $\theta_{\rm AS} <45\arcsec$.  For the polarised ERS resolved in NVSS we find that the WISE--AGN polarised ERS have a median angular size of $55\pm1\arcsec$ and a median percentage polarisation of $3.6\pm0.2\,$\%, while the WISE--Star polarised ERS are larger with a median angular size of $66\pm3\arcsec$ and a median percentage polarisation of $6.3\pm0.9\,$\%.  For our sources that are unresolved in NVSS we find that the WISE--AGN polarised ERS have a median angular size of $31\pm1\arcsec$ and a median percentage polarisation of $3.3\pm0.1\,$\%, while the WISE--Star polarised ERS are larger with a median angular size of $35\pm1\arcsec$ and a median percentage polarisation of $9.5\pm1.4\,$\%.  The percentage polarisation for the WISE--AGN and WISE--Star populations is constant for the two angular size bins.

In Fig.~\ref{Lsize} we plot the linear size of both classifications of polarised ERS as a function of percentage polarisation and total flux density.  We calculated the linear size of the polarised ERS using the method outlined by \citet{Hogg1999}.  The polarised WISE--AGN ERS population tends to be larger in linear size compared to the polarised WISE--Star ERS population.  We have also plotted the median linear size values for the WISE--AGN population (black dots) and the WISE--Star (red stars) populations.  Our two polarised ERS populations show a large difference in linear size.  The median linear size of the WISE--AGN population is $\theta_{\rm LS} = 253\,\pm\,8\,$kpc and  for the WISE--Star population $69\,\pm\,8\,$kpc.  We note that the median linear size value remains constant for both populations as a function of both percentage polarisation and total flux density.

%-------------------------------------------------------------------------
%                         SELECTION EFFECTS
%-------------------------------------------------------------------------
\section{Selection Effects}\label{sec:seleff}
We acknowledge that there are selection effects with our polarised ERS populations and we discuss each briefly.
\vskip 0.2cm
\noindent 1.  WISE detections: WISE was built to survey the entire sky in the mid-infrared.  As a result, WISE will observe the full extent of the obscured AGN and QSO and detect the Ultra-luminous Infrared Galaxies \citep{Wright2010}.  However, WISE is not sensitive to elliptical or lenticular galaxies at high redshifts as these types of galaxies contain little dust and gas and will fall below the detection threshold of WISE.  Elliptical galaxies are known to host powerful AGN so our sample of polarised WISE--Star sources is biased toward low redshift, whereas our sample of WISE--AGN sources will be detectable out to $z = 3$ \citep{Wright2010}.  We also note that highly luminous quasar-like galaxies are rare at low redshift and as a result of these effects our two populations of polarised ERS do not overlap significantly in redshift space, see Fig.~\ref{redshiftdist}.
\vskip 0.2cm
\noindent 2.  Redshift selection: The \citet{Hammond2012} catalogue contains redshifts from various optical surveys with different sensitivity limits as mentioned in Section \ref{sec:RMcat}.  A number of high-redshift AGN will not be detected in these surveys as the optical host galaxy is fainter than their counterpart at low-redshift.  Therefore, the \citet{Hammond2012} catalogue will be biased towards optically brighter and more nearby AGN.
\vskip 0.2cm
\noindent 3.  Polarised flux density detection limit:  In all polarisation studies there is a bias towards high percentage polarisation near the detection limit of the images.  We see this bias in our sample (Fig.~\ref{lum}) where there is a shortage of polarised ERS with $\Pi \le 3\,$\% at $L_{\rm 1.4\,GHz} \le 10^{25}\,$W Hz$^{-1}$.  Both the WISE--AGN and WISE--Star ERS suffer from selection effects in the same way.  There are no highly polarised WISE--AGN at $S_{\rm 1.4\,GHz} \le 100\,$mJy in Fig.~\ref{logIlogP}(a) which, if there are any, should have been detected.  Instead, we only detect highly polarised WISE--Star ERS at these flux density levels.  
\vskip 0.2cm
Although WISE is sensitive to certain types of galaxies, we find that our sample splits nicely into two separate host galaxy populations as shown in Fig.~\ref{colormag} and with this split we analyse the properties of both samples.  We acknowledge that our sample is not statistically complete and that we may be affected by unknown selection effects not mentioned above.  We notice this possibility in the fact that our mean $\Pi$ for both populations is higher than estimated by both \citet{Mesa2002} and \citet{Tucci2004}.  We also notice in both Fig.~\ref{logIlogP}(a) and Fig.~\ref{lum} that there is a lack of highly polarised WISE--AGN sources across all flux densities whereas the WISE--Star population begins to fill out this area at $S_{\rm 1.4\,GHz} \le 100\,$mJy regardless of the fact that only highly polarised sources are detected at the faint flux density levels.  The fact that we do not detect the WISE--AGN with high percentage polarization at $S_{\rm 1.4\,GHz} \le 100\,$mJy demonstrates a change in the intrinsic properties of polarised ERS as the percentage polarization is greater for WISE--Star than for WISE--AGN at these flux densities.  In the next section we discuss possible astrophysical reasons for our findings noting these selection effects in our data.

%-------------------------------------------------------------------------
%                                DISCUSSION
%-------------------------------------------------------------------------
\section{Discussion}\label{sec:dis}
\subsection{Is the polarisation correlation with flux real?}\label{sec:flux}
\citet{Mesa2002} and \citet{Tucci2004} found an inverse correlation between the percentage linear polarisation and total flux densities of NVSS sources, so that faint sources were more highly polarised. A similar result was found for the ELAIS-N1 sources by \citet{Taylor2007} and \citet{Grant2010}, and for ATLBS sources by \citet{Subrahmanyan2010}.  \citet{Rudnick2014} shows that the \citet{Grant2010} completeness correction is too large at the faintest polarised flux density bins but they also come to the conclusion that the population of polarized radio sources changes in composition.

However, \citet{Hales2013}, while finding an observational increase in percentage polarisation with decreasing flux density in the ATLAS data set, attributed this entirely to selection effects, including the non-detections which must be accounted for in a full statistical treatment. Once the data were corrected for these effects, \citet{Hales2013} found that the percentage polarisation of their sources showed no dependence on flux density, agreeing with results at higher frequencies by \citet{Massardi2013}. These results cast doubt on earlier results that had found such dependence.

We show that the correlation between percentage polarisation and flux density is real and can not be the result of selection effects. Fig.~\ref{logIlogP} shows that sources with a WISE classification of ÒstarÓ (i.e. passive and quiescent galaxies) show an increase in fractional polarisation with decreasing flux density, while a weaker trend is found for sources with a WISE classification of ÒAGNÓ (i.e. luminous quasar-like galaxies).  Selection effects in total flux density affect both the WISE--Star and WISE--AGN populations at low total flux density and polarised flux density.  If the overall increase with percentage polarisation with decreasing flux density were due to bias or selection effects in the radio data, then, for a given radio flux density or percentage polarisation, these effects would show no correlation with WISE colours. This is inconsistent with Fig.~\ref{colormag-pi} and Fig.~\ref{logIlogP}, which demonstrates that the percentage polarisation depends on WISE colours.

Given that the effect is demonstrably present in our data set, we can then examine why \citet{Hales2013} failed to detect it after removing selection effects. Data from \citet{Hales2013} probe much lower radio flux densities, so probe much larger redshift range, over a much smaller area than our data set. Since quiescent galaxies result from hierarchical merging of star-forming galaxies, they are less numerous at high redshift than at low redshift. It is therefore likely that the data from \citet{Hales2013} include a smaller number of WISE--Star galaxies, and so the trend we see here will be very much reduced in the \citet{Hales2013} data set.  If this explanation is correct, then the inconsistency between \citet{Hales2013} and other authors is an indication not of problems with the data, but of cosmic evolution of the space density of quiescent galaxies. This will be investigated further in the ATLAS Data Release 3 by Banfield et al. (in preparation).

\subsection{The origin of the polarisation correlation with flux}\label{sec:origin}
\citet{Shi2010} and \citet{Banfield2011} conclude that the inverse correlation of percentage polarisation and total flux density may be a result of the intrinsic properties of the polarised ERS.  Using ERS with $\Pi > 30\,$ per cent, \citet{Shi2010} show that polarised ERS at $1.4\,$GHz are contained within elliptical galaxies and that there is no dependence on the source environment compared to low-polarisation ERS.  \citet{Banfield2011} confirmed that there is no trend of percentage polarisation with redshift, however, a trend of increasing percentage polarisation with decreasing luminosity was found.  \citet{Hammond2012} found that polarised ERS with optical counterparts classified as galaxies have higher polarisation percentages compared to polarised ERS with optical classifications as quasars, agreeing that a population change in polarised ERS may be the cause of the inverse correlation between percentage polarisation and total flux density.  

Our sample of polarised ERS shows a strong distinction between two different galaxy populations.  The two populations differ in both infrared colours and radio polarisation properties.  The WISE--AGN ERS are radio-loud AGN at high radio luminosity and have larger linear sizes compared to the WISE--Star ERS.  The infrared emission from the WISE--AGN can originate from dust that is being heated by some combinations of AGN and star formation activity.   The WISE--Star ERS are also radio-loud AGN at lower radio luminosity and have smaller linear sizes.  The infrared emission from the WISE--Star originates from the stars within the galaxy, pointing to an old elliptical galaxy.  The WISE--AGN population show lower percentage polarisation than the WISE--Star population.

Our results confirm the previous \citet{Taylor2007}, \citet{Banfield2011}, and \citet{Hammond2012} that the percentage polarisation increases with decreasing total flux density is the result of a population change.  Our sample of highly polarised ERS are found to be part of the WISE--Star population, endorsing the result from \citet{Shi2010} that the polarised radio sources are inside elliptical galaxies.  Since \citet{Hales2013} find no such trend after removing selection effects, we conclude that the \citet{Hales2013} data set contain fewer WISE--Star ERS. Further investigation is required to determine if there is an evolution of one population of polarised ERS to another population of polarised ERS.

\subsection{Environments of polarised radio sources}\label{sec:environ}
High-redshift radio galaxies \citep[HzRGs;][]{Seymour2007,DeBreuck2010} are radio galaxies found at $1<z<5$ and are indicators of large overdensities or proto-clusters in the early Universe \citep{Miley2008}.  Our sample of polarised WISE--AGN lie between $z=0.5$ and $z=3.7$ with luminosities in the range of $9 \times 10^{23} < L_{\rm 1.4} < 7 \times 10^{28}\,$W Hz$^{-1}$; placing our sources in the category of HzRGs. \citet{Humphrey2013} provide evidence that HzRGs in these over dense regions can be surrounded by giant ionised gas halos.    Radio observations of Cygnus A by \citet{Dreher1987} revealed large fluctuations in RM across the lobes and magnetic field reversals of the order $\sim$20 kpc.  \citet{Dreher1987} suggest that the intracluster medium (ICM) or a sheath surrounding the radio lobes can cause these high RMs.   Models by \citet{Bicknell1990} suggest a turbulent interface between the magnetised plasma in the radio lobe and the ICM causing reversals in the magnetic field.  Recent observations of Centaurus A by \citet{Osul2013} provide evidence of depolarisation across radio lobes from the presence of a significant amount of thermal gas within the lobes.  \citet{Bell2013} examine a sample of radio sources exhibiting the Laing-Garrington effect and show that the depolarisation cannot be explained by beaming.  \citet{Farnes2014} compared total intensity spectral indices with polarised spectral indices to show that there are two populations of polarised radio sources: core- and jet-dominated sources.  \citet{Farnes2014} suggest that these two different source populations undergo different depolarisation mechanisms based on the local source environment.  Our work support the conclusions from other authors that WISE--AGN have larger linear sizes consistent with being more powerful and lower percentage polarisation.  As the radio galaxy expand through the over dense regions surrounding the host galaxy, mixing between the radio lobes and the surrounding gas tangles the magnetic field line, which results in depolarisation.

\subsection{Extragalactic rotation measures}\label{sec:erms}
The WISE--Star and WISE--AGN populations have different RM variances, which is a combination of the different physical properties of these radio sources (star-forming galaxies versus AGN) and from the longer lines of sight towards the WISE--AGN sources, which can host more intervening objects that leave their imprint on the RMs of the background sources.
We write the extragalactic RM contributions that we find after removing the Galactic foreground contribution as
\begin{eqnarray}\label{eqn:rm_extragalactic}
\mathrm{RM} & = & \mathrm{RM}_\mathrm{MW,res} + \mathrm{RM}_\mathrm{web} + \mathrm{RM}_\mathrm{cluster} + \mathrm{RM}_\mathrm{Mg\,{II}} +  \nonumber \\
& & \frac{\mathrm{RM}_\mathrm{int}}{(1+z)^2} + \mathrm{rest} \, , 
\end{eqnarray}
\noindent
which decomposes the extragalactic RM into contributions by the Milky Way that have not been completely removed, intervening structures in the cosmic web, galaxy clusters, and Mg\,{II} absorbers, the source-intrinsic RMs corrected for their redshift $z$, and a rest term which we will show plays only a very small role. Assuming that the RMs in Equation~\ref{eqn:rm_extragalactic} follow Gaussian distributions, the RM variances can be written as:
\begin{eqnarray}\label{eqn:rm_variance_extragalactic}
\sigma_\mathrm{RM}^2 & = & \sigma_\mathrm{RM,MW_\mathrm{res}}^2 + \sigma_\mathrm{RM,web}^2 + \sigma_\mathrm{RM,cluster}^2 + \sigma_\mathrm{RM,Mg\,{II}}^2 + \nonumber \\
& & \frac{\sigma_\mathrm{RM,int}^2}{\left(1+z\right)^4} + \sigma_\mathrm{rest}^2 \, .
\end{eqnarray}
\noindent
Here we used $\langle \mathrm{RM}_\mathrm{MW,res}\rangle \approx 0$~rad m$^{-2}$ from our simulations, and the fact that the net RMs in intervening objects are zero on average. While clusters can have a considerable impact on the RM variance of background sources (e.g., \citealt{Clarke2004}), \citet{Johnston-Hollitt2011} showed that the mean RM of sources from the NVSS catalogue that lie behind clusters is $\approx$ 0~rad m$^{-2}$.

From the simulations by \citet{Akahori2011} and the observations by \citet{Joshi2013} we can estimate the contributions by cosmological large-scale structure and intervening Mg\,{II} absorbers, respectively, on the extragalactic RM variances that we observed.  While some of Mg\,{II} absorption can be associated with the host quasars \citep[e.g.,][]{Farina2014} or host galaxies \citep[e.g.,][]{Bordoloi2014} themselves, in our analysis we should only correct for the contribution by intervening Mg\,{II} absorbers that are not associated with the host quasars or galaxies.
In their study of the RM imprint of intervening Mg~II absorbers, \citet{Joshi2013} considered only Mg\,{II} absorbers with relative velocities of more than 5000 km~s$^{-1}$ with respect to the background quasars.
We estimate the contribution by intervening Mg\,{II} absorbers to the $\sigma_\mathrm{RM}$ we derived based on this analysis by \citet{Joshi2013} of intervening Mg\,{II} systems.

For the high-redshift WISE--AGN (median redshift of 1.02) large-scale structure contributes 7--8~rad m$^{-2}$. 
Models `ALL' and `CLS' from \citet{Akahori2011} predict larger standard deviations in RM, but contain contributions by galaxy clusters that could not be properly modeled given the cell size of the simulations, as the authors mention in their section~2.2.
\citet{Joshi2013} found that sightlines with Mg\,{II} absorbers show an increase in the standard deviations in RM by 8.1 $\pm$ 4.8~rad m$^{-2}$, and that about 1/3 of the sightlines towards high-redshift sources contain one or more Mg\,{II} absorbers. Combining the increased RM variance due to intervening Mg\,{II} absorbers with the frequency with which such systems are encountered gives $\sigma_\mathrm{RM,Mg\,{II}}^2$~ =~18.4~$(\mathrm{rad\ m}^{-2})^2$. 
For the low-redshift WISE--Star sources (median redshift of 0.06) large-scale structure contributes $\approx$ 1.4~rad m$^{-2}$, and we assume that the contribution by Mg\,{II} absorbers is negligible given the short sightlines towards the WISE--Star sources.

With this information we can write out Equation~\ref{eqn:rm_variance_extragalactic} separately for the low-redshift WISE--Star sources and the high-redshift WISE--AGNs, and subtract the two expressions. Using the RM variances that we derived for the WISE--Star and WISE--AGN samples that we determined in Section~\ref{sec:rm}, one can show that
\begin{eqnarray}\label{eqn:rm_extragalactic_balance}
\lefteqn{ 
\left(\sigma_\mathrm{RM,cluster}^2 + \frac{\sigma_\mathrm{RM,int}^2}{\left(1+z\right)^4}\right)_{z\approx 1} =  
} \nonumber \\
& & \left(\sigma_\mathrm{RM,cluster}^2 + \frac{\sigma_\mathrm{RM,int}^2}{\left(1\right)^2}\right)_{z\approx 0} + 8 \, (\mathrm{rad~m}^{-2})^2 \, ,
\end{eqnarray}
where the numerical term was calculated as 144-64-18-(56-2) $(\mathrm{rad~m}^{-2})^2$, combining the extragalactic RM variances of the high-redshift WISE--AGNs and low-redshift WISE--Star sources, and the contribution by the cosmic web and Mg\,{II} absorbers. Given the very different physical properties of the WISE--AGN and WISE--Star sources, which are reflected in $\sigma_\mathrm{RM,int}$, and the contributions by clusters that we could not estimate, the `excess' of 8 $(\mathrm{rad\ m}^{-2})^2$ in the variance is very small. This excess includes the difference in residual contributions by the Milky Way for the WISE--AGN and WISE--Star sources, and contrbutions by sources along the line of sight that we included as the rest term in Equations~\ref{eqn:rm_extragalactic} and \ref{eqn:rm_variance_extragalactic}. 

Long lines of sight have a higher chance of passing through galaxy clusters than short lines of sight, and as a result, $\left(\sigma_\mathrm{RM,cluster}^2\right) _{z\approx 1} > \left(\sigma_\mathrm{RM,cluster}^2\right) _{z\approx 0}$. Therefore Equation~\ref{eqn:rm_extragalactic_balance} also implies that the distribution of the source-intrinsic RMs of the WISE--AGN, measured in terms of the standard deviation, is at most four times as wide as the distribution of RMs of the WISE--Star star-forming galaxies, if the standard deviation of the WISE-Star galaxies itself is not too small.

%%% CONCLUSION %%%
\section{Conclusions}\label{sec:conc}
Using the \citet{Hammond2012} catalogue of Faraday rotation measures and redshifts for 4003 ERS detected at $1.4\,$GHz, we have shown that polarised radio sources split into two types of host galaxies at two separate redshift ranges, as such the two populations are investigated separately and a larger sample of polarised radio sources is required to examine if one population evolves into the other population.  We find the following:
\vskip 0.2cm
\noindent (1) the anti-correlation between percentage polarisation and total flux density is real as the percentage polarisation depends on WISE mid-infrared colour;
\vskip 0.2cm
\noindent (2) the polarised ERS separate clearly into two infrared-selected objects: WISE--Star sources that are low-redshift, low-radio-luminosity elliptical galaxies, and WISE--AGN which are high-redshift, high-radio-luminosity quasar-like galaxies; 
\vskip 0.2cm
\noindent (3) our sample has a larger number of quiescent galaxies than \citet{Hales2013}, suggesting that the inconsistency between the data sets is an indication of cosmic evolution of the space density of quiescent galaxies;
\vskip 0.2cm
\noindent (4) we suggest that the difference in the percentage polarisation of radio galaxies originates from the environment of the host galaxy.  Our WISE--AGN population is consistent with HzRGs in denser environments where depolarisation is more severe compared to the WISE--Star sources that are not very active;
\vskip 0.2cm
\noindent (5) we find that the extragalactic RM contributions to the nearby WISE--Star and the distant WISE--AGN sources are different; the distribution of source-intrinsic RMs of the WISE--AGNs is at most four times as wide as the distribution of intrinsic RMs of the star-forming WISE--Star galaxies if the distribution of intrinsic RMs of the WISE--Star sources itself is at least several rad m$^{-2}$ wide; and
\vskip 0.2cm
\noindent (6) we also detect no evolution of RM with redshift, suggesting that the RM is a product of the intrinsic properties of the radio galaxy and not a result of the intervening large-scale structure of the Universe.

\section*{Acknowledgments}
The authors would like to thank H-R. Kl\"ockner for his useful comments on the manuscript, B.~M.~Gaensler for his advice on the rotation measure catalogue, L. Rudnick for the discussion on rotation measures and polarised source populations, and the anonymous referee for the suggestions making our paper stronger.  This publication makes use of data products from the Wide-field Infrared Survey Explorer, which is a joint project of the University of California, Los Angeles, and the Jet Propulsion Laboratory/California Institute of Technology, funded by the National Aeronautics and Space Administration.  This publication makes use of data products from the Sloan Digital Sky Survey.  Funding for the SDSS and SDSS-II has been provided by the Alfred P. Sloan Foundation, the Participating Institutions, the National Science Foundation, the U.S. Department of Energy, the National Aeronautics and Space Administration, the Japanese Monbukagakusho, the Max Planck Society, and the Higher Education Funding Council for England. The SDSS Web Site is http://www.sdss.org. 

\newcommand{\aj}{AJ} 
\newcommand{\apj}{ApJ}
\newcommand{\mnras}{MNRAS} 
\newcommand{\pasa}{PASA}
\newcommand{\aap}{A\&A}
\newcommand{\araa}{ARA\&A}
\newcommand{\aaps}{A\&AS}
\newcommand{\aujpa}{AuJPA}
\newcommand{\apjs}{ApJS}
\newcommand{\newar}{New Astron.~Rev.}
\newcommand{\aar}{A\&ARv}
\newcommand{\apss}{Ap\&SS}
\newcommand{\japa}{JAp\&A}
\newcommand{\jkas}{JKAS}

\bibliographystyle{mn2e}

\appendix

\section[]{Derivation of extragalactic rotation measures}\label{sec:appendixA}
Consider a population of polarised extragalactic radio sources where each source has a rotation measure (RM).  This RM contains contributions from all along the sight: the intrinsic RMs of the sources themselves; intergalactic space; the Milky Way; and the Earth's ionosphere. The RMs that are built up inside and outside the Milky Way are much larger than the ionospheric RM \citep{Sotomayor2013}. Using the RM values from the catalogue by \citet{Taylor2009}, \citet{Schnitzeler2010} has shown that the RM contribution by the Galactic foreground can be separated from the contribution from outside the Milky Way because the foreground RM contributions are correlated between sightlines, while the latter does not depend on the viewing direction in the Milky Way. Here we describe a simple method to calculate the extragalactic RM variance of the radio sources. First we analyse how the RMs from different parts of the line of sight contribute to the RM variance that we measure for an ensemble of extragalactic sources, then we outline a five-step method to calculate the extragalactic RM variance of this ensemble. Similar to \citet{Schnitzeler2010} we only include sightlines that lie further than 20$^\circ$ from the Galactic plane.

We use cubic spline fitting as described in \citet{Schnitzeler2010} to remove large-scale RM gradients along Galactic longitude that are induced by the Milky Way. All RMs from the catalogue by \citet{Taylor2009} are included in this cubic spline fitting, also when we analyse the WISE--AGN and WISE--Star subpopulations. The strips that we use to fit and remove a cubic spline are four degrees wide in Galactic latitude. We also calculate and remove a cubic spline fit to strips that are shifted by two degrees (half a strip width) to provide Nyquist sampling of all Galactic latitudes. 

The observed RM of a radio source can be split into a contribution by the Milky Way, a contribution from outside the Milky Way, and the measurement error in RM as:
\begin{equation}\label{eqn:A1}
\mathrm{RM}_\mathrm{obs} = \mathrm{RM}_\mathrm{MW} + \mathrm{RM}_\mathrm{ERS} + \mathrm{RM}_\mathrm{err} \, .
\end{equation}
After cubic spline fitting, the variance of an ensemble of sightlines in a single strip along Galactic longitude can be written as:
\begin{equation}\label{eqn:A2}
\sigma_\mathrm{RM}^2 = \sigma_\mathrm{MW}^2 + \sigma_\mathrm{ERS}^2 + \sigma_\mathrm{err}^2 \, ,
\end{equation}
since the contributing RMs all have the same mean value (of 0 rad~m$^{-2}$).
This ensemble can be a subset from the catalogue by \citet{Taylor2009}. When multiple strips are combined the variance of the ensemble of sightlines can be written as:
\begin{eqnarray}\label{eqn:A3}
\sigma_\mathrm{RM}^2 & = & \frac{1}{N_\mathrm{los}-1}\sum_{i=1}^{N_\mathrm{los}}(\mathrm{RM}_\mathrm{obs} - \langle\mathrm{RM}\rangle_{1+...+M})^2 \nonumber \\ 
& = &  \frac{N_1 - 1}{N_\mathrm{los}-1}\frac{1}{N_1-1}\sum_{\mathrm{strip}\ 1}(\mathrm{RM}_\mathrm{obs} - \langle\mathrm{RM}\rangle_\mathrm{strip\ 1})^2 \nonumber\\ 
& &  + ... + \nonumber\\ 
&  & \frac{N_M - 1}{N_\mathrm{los}-1}\frac{1}{N_M-1}\sum_{\mathrm{strip}\ M}(\mathrm{RM}_\mathrm{obs} - \langle\mathrm{RM}\rangle_{\mathrm{strip}\ M}\rangle)^2 + \nonumber \\
& & \frac{N_1}{N_\mathrm{los}-1}\left(\langle\mathrm{RM}\rangle_\mathrm{strip\ 1} - \langle\mathrm{RM}\rangle_{1+...+M}\right)^2 \nonumber\\
& &  + ... + \nonumber\\ 
& & \frac{N_M}{N_\mathrm{los}-1}\left(\langle\mathrm{RM}\rangle_{\mathrm{strip}\ M} - \langle\mathrm{RM}\rangle_{1+...+M}\right)^2, 
\end{eqnarray}
\noindent
if $N_\mathrm{los}$ sightlines are distributed over $M$ strips. $N_\mathrm{los}$ = $N_1 + ... + N_{M}$, where $N_\mathrm{i}$ is the number of sightlines in strip $i$.
$\langle\mathrm{RM}\rangle_{1+...+M}$ and $\langle\mathrm{RM}\rangle_{\mathrm{strip}\ i}$ indicate the mean RM of the ensemble of all strips and the mean RM of all sightlines in a single strip, respectively.
Using Equation~\ref{eqn:A2} to re-write Equation~\ref{eqn:A3}, and combining the $\sigma_\mathrm{ERS}^2$ and $\sigma_\mathrm{err}^2$ terms from the different strips, Equation~\ref{eqn:A3} can be written as (equation \ref{eqn:rm} in text):
\begin{eqnarray}\label{eqn:A4}
\sigma_\mathrm{RM}^2 & = & \left(\sigma_\mathrm{ERS}^2 + \sigma_\mathrm{err}^2\right)\left(\frac{N_\mathrm{los}-N_\mathrm{strips}}{N_\mathrm{los}-1}\right) + \nonumber \\
& &  \frac{N_1 - 1}{N_\mathrm{los}-1}\sigma_\mathrm{MW,1}^2 + ... + \frac{N_M - 1}{N_\mathrm{los}-1}\sigma_{\mathrm{MW},M}^2 + \nonumber\\
& & \frac{N_1}{N_\mathrm{los}-1}\left(\langle\mathrm{RM}\rangle_\mathrm{strip\ 1} - \langle\mathrm{RM}\rangle_{1+...+M} \right)^2 \nonumber\\
& &  + ... + \nonumber\\ 
& & \frac{N_M}{N_\mathrm{los}-1}\left(\langle\mathrm{RM}\rangle_{\mathrm{strip}\ M} - \langle\mathrm{RM}\rangle_{1+...+M} \right)^2, 
\end{eqnarray}
if there are $N_\mathrm{strips}$ with usable sightlines.

Using Equation~\ref{eqn:A4} we can calculate the variance of the extragalactic RMs of the radio sources in the following way:\\
1. Calculate $\sigma_\mathrm{RM}^2$ for the ensemble of all sightlines after large-scale RM gradients have been removed by cubic spline fitting.\\
Then (working from right to left in Equation~\ref{eqn:A4}):\\
2. Correct for the difference between the mean RM of the ensemble of all sightlines and the mean RM of sightlines belonging to  a single strip,\\
3. Subtract the contribution by the Milky Way, \\
4. Divide by the bias-correction term, and\\
5. Subtract the variance in RM that is expected purely due to the measurement errors of the RMs.

The mean RM of individual strips and of the ensemble of all sightlines were found using robust statistics, where RM outliers at the 3-sigma level were removed from the ensemble.

\citet{Schnitzeler2010} found that the Milky Way contributes $\sigma_\mathrm{RM,MW}$ = 6.8 rad m$^{-2}$/$\sin\left(\mathrm{latitude}\right)$ at positive Galactic latitudes and $\sigma_\mathrm{RM,MW}$ = 8.4 rad m$^{-2}$/$\sin\left(-\mathrm{latitude}\right)$ at negative Galactic latitudes. To correct for how these contributions change with Galactic latitude we bin the ensemble of sightlines into strips along Galactic longitude; the width of these strips does not have to be the same as the width that we used to remove large-scale RM gradients by cubic spline fitting. In our analysis we only used sightlines from a strip if the number of sightlines in that strip is larger than a threshold value. We varied both the strip width and the threshold value to check how robust our results are. 

The bias-correction term $\left(N_\mathrm{los}-N_\mathrm{strips}\right)/\left(N_\mathrm{los}-1\right)$ combines the bias-correction terms of the individual strips. 
The remaining distribution of RMs consists of a contribution by the RMs that are built up outside the Milky Way, and a contribution by the measurement errors in RM. We use a Monte Carlo process to simulate the width of the distribution if there is only noise from the measurement errors, and no astrophysical signal. For each sightline we draw an RM from a Gaussian distribution with zero mean and a standard deviation that is equal to the measurement error in RM of that sightline. We then determine the standard deviation and variance of the RM distribution of the ensemble of sightlines, and repeat this process 5000 times to build up a distribution of standard deviations and variances. We use the square of the mean standard deviation of these 5000 runs for $\sigma_\mathrm{err}^2$; using the mean of the variances gives negligible differences for the $\sigma_\mathrm{ERS}^2$ that we derive.

We estimate the uncertainty in $\sigma_\mathrm{ERS}$ in Equation~\ref{eqn:A4}, $\mathrm{err}\left(\sigma_\mathrm{ERS}\right)$, from the uncertainties in $\sigma_\mathrm{err}^2$, $\sigma_{\mathrm{MW},i}^2$,$ \langle\mathrm{RM}\rangle_{\mathrm{strip}\ i}$, and $\langle\mathrm{RM}\rangle_{1+...+M}$ using the standard expression for error propagation. 
We did not include the error in $\sigma_\mathrm{RM}^2$ because this is difficult to estimate; the error in $\sigma_\mathrm{ERS}$ that we derive should therefore be interpreted as a lower limit on the true error. 
The error in $\sigma_\mathrm{err}^2$ depends on whether one uses $\sigma_\mathrm{err}^2 = \langle\sigma_\mathrm{ERS}\rangle^2$ or $\sigma_\mathrm{err}^2 = \langle\sigma_\mathrm{ERS}^2\rangle$, but in practice these error terms contribute little to the overall error in $\sigma_\mathrm{ERS}$ for the ensemble of all sightlines, and for the subsamples of sightlines towards WISE--Star and WISE--AGN sources.
In the first case $\mathrm{err}\left(\langle\sigma_\mathrm{err}\rangle^2\right) = \mathrm{SD}\left(\sigma_\mathrm{err}\right)/\sqrt{N_\mathrm{MC}}$, where $\mathrm{SD}\left(\right)$ calculates the standard deviation of the argument, and $N_\mathrm{MC}$ indicates the number of Monte Carlo runs. 
In the second case $\mathrm{err}\left(\langle\sigma_\mathrm{err}^2\rangle\right)= \mathrm{SD}\left(\sigma_\mathrm{err}^2\right)/\sqrt{N_\mathrm{MC}}$.

The other errors are easier to calculate. $\mathrm{err}\left(\sigma_{\mathrm{MW},i}^2\right)$ can be derived from $\mathrm{err}\left(\sigma_{\mathrm{MW},i}\right) \lesssim 0.5\ \mathrm{rad\ m}^{-2}$ based on how sensitive the model curves in figure~3 from \citet{Schnitzeler2010} are to even small changes in $\sigma_\mathrm{MW}$; $\sigma_\mathrm{MW} = 0.5\ \mathrm{rad\ m}^{-2}$ is a conservative upper limit. 
$\mathrm{err}\left(\langle \mathrm{RM}_{\mathrm{strip}\ i}\rangle\right)$ and $\mathrm{err}\left(\langle \mathrm{RM}_{1+...+M}\rangle\right)$ 
can be derived from $\mathrm{SD}\left(\mathrm{RM}_{\mathrm{strip}\ i}\right)/\sqrt{N_i}$ and from $\mathrm{SD}\left(\mathrm{RM}_{1+...+M}\right)/\sqrt{N_\mathrm{los}}$, respectively.

\bsp

\label{lastpage}

\end{document}